\documentstyle[aps,prl,multicol]{revtex}

\newcommand{\QKD}{{\sc qkd}}
\newcommand{\PDC}{{\sc pdc}}
\newcommand{\WCP}{{\sc wcp}}
\newcommand{\ra}{\rangle}
\newcommand{\la}{\langle}
\begin{document}
\title{Security Aspects of Practical Quantum Cryptography}
\author{Gilles Brassard$^{1}$, Norbert L\"utkenhaus$^{2}$, Tal Mor$^{1,3}$
and Barry C.\ Sanders$^{4}$}

\address{$^1$ D\'epartement IRO, Universit\'e de Montr\'eal,
C.P. 6128, succ.~centre--ville, Montr\'eal (Qu\'ebec), Canada H3C 3J7\\
{$^2$}Helsinki Institute of Physics, P.O.Box 9, 00014 Helsingin
yliopisto, Finland\\
{$^3$}Electrical Engineering, University of California at Los Angeles,
Los Angeles, CA 90095--1594, USA\\
{$^4$}Department of Physics, Macquarie University,
Sydney, New South Wales 2109, Australia \\
}

\date{\today}
\maketitle
\begin{abstract}

The use of quantum bits (qubits) in cryptography holds the promise of
secure cryptographic quantum key distribution schemes.  Unfortunately,
the implemented schemes can be totally insecure.  We provide a
thorough investigation of security issues for practical quantum key
distribution, taking into account channel losses, a realistic
detection process, and modifications of the ``qubits'' sent from the
sender to the receiver.  We first show that even quantum key
distribution with perfect qubits cannot be achieved over long
distances when fixed channel losses and fixed dark count errors are
taken into account.  Then we show that existing experimental schemes
(based on ``weak-pulse'') are usually totally insecure.  Finally we
show that parametric downconversion offers enhanced performance
compared to its weak coherent pulse counterpart.

\end{abstract}
\vspace{5mm}
Pacs:{ 03.67.Dd, 42.50.Dv, 03.65.Bz, 89.80.+h}

\begin{multicols}{2}

Quantum information theory suggests the possibility of accomplishing
tasks which are beyond the capability of classical computer science,
such as information-secure cryptographic key
distribution~\cite{BB84,BBE}.
The lack of security proofs for standard (secret- and public-)
key distribution schemes, and the insecurity of the strongest classical
schemes against ``quantum attacks'' \cite{Shor97},
emphasizes the need for
information-secure key distribution.
Whereas the security of idealized
quantum key distribution (\QKD) schemes has been investigated
against very sophisticated collective and 
joint attacks (e.g.,~\cite{BM97b,mayers}), the
experimental \QKD\ schemes have been proven secure against the
simple individual attack only recently~\cite{nl99b}
(via the application of ideas presented here). 

In the four-state scheme~\cite{BB84}, usually referred to as
Bennett-Brassard-84 (BB84), the
sender (Alice) and the receiver (Bob) use two conjugate bases (say,
the rectilinear basis, $+$, and the diagonal basis, $\times$) for the
polarization of single photons.  In basis $+$ they use the two
orthogonal basis states $|0_+ \rangle$ and $|1_+ \rangle $ to
represent ``0'' and ``1'' respectively.  In basis $\times$ they use
the two orthogonal basis states $|0_\times \rangle = (1/\sqrt2) [|0_+
\rangle + |1_+ \rangle] $ and $|1_\times \rangle = (1/\sqrt2) [|0_+
\rangle - |1_+ \rangle] $ to represent ``0'' and ``1''.  The basis is
revealed later on via an unjammable and insecure classical
channel. The signals where Bob used the same basis as Alice form the
{\em sifted key} on which Bob can decode the bit value. The remaining
signals are being discarded.
Finally, they test 
a few bits to estimate the error-rate, and if the test
passes (the tested error-rate is less than some pre-agreed threshold),
they use error-correction and privacy amplification to obtain
a potentially secure final key~\cite{BBBSS,bennett95a}.

The security of that scheme, which assumes a source of perfect qubits
as well as losses and errors which are bounded by some small threshold,
has been investigated in various works.
Very simple attacks already render realistic \QKD\
impossible, as we show here. 

The experiments are usually based on
weak coherent pulses (\WCP)
as signal states with a low probability of
containing more than one photon \cite{BBBSS,WCPexperiments}.
Initial security analysis of such weak-pulse schemes
were done~\cite{BBBSS,HIGM}, and evidence of some potentially severe
security problems (which do not exist for the idealized schemes)
were shown~\cite{HIGM,Yuen}. 
We investigate such limitations much further
to show insecurity of various existing setups, 
and to provide several explicit limits on
experimental \QKD.

First, we show that \QKD\ 
can be totally impossible for given losses and detector dark-counts,
even with the assumption of a
perfect source.
Second 
we show that \QKD\
can be totally insecure even 
with the assumption of perfect detection, if considering  losses and  multi-photon states.
In a combination we conclude that,
for any given source and detection units,
secure \QKD\ schemes cannot be implemented 
to arbitrarily large distance.
We  analyse 
``weak pulse'' schemes and show that the implemented schemes
are insecure.
Finally we prove the advantage
of a better source  such as parametric downconversion,
and we show that it presents characteristics very
similar to the semi-idealized schemes where perfect qubits are assumed.

The effect of losses is that the signals will arrive only
with a probability $F$ at Bob's site where they will lead to a
detection in Bob's detectors with a probability $\eta_{\rm B}$ (detection
efficiency). This leads to an expected probability of detected signals
given by $p_{\rm exp}^{\rm signal} = F \eta_{\rm B}$. The transmission efficiency $F$ is
connected to the absorption coefficient $\beta$ of the fibre, the
length $l$ of the fibre and a distance-independent constant loss in
optical components $c$, via the relation
\begin{equation}
\label{distance}
F = 10^{-\frac{\beta \; l + c}{10}}
\end{equation} which, for given $\beta$ and $c$,
gives a one-to-one relation between distance and transmission
efficiency.  
Bob's detector is also
characterized by a dark count
probability $p_{\rm exp}^{\rm dark}=d_{\rm B}$.
The dark counts are due to thermal fluctuations in the detector, stray
counts, etc. 
In practice, $d_B$ 
is inferred
as counts per time slot in the absence of
the real signal. The total expected probability of detection events is
therefore given by
$p_{\rm exp} = p_{\rm exp}^{\rm signal} + 
p_{\rm exp}^{\rm dark}-p_{\rm exp}^{\rm signal}
p_{\rm exp}^{\rm dark}$ 
and neglecting the coincidence term (all involved probabilities are
small) we get
\begin{equation} \label{p-exp}
p_{\rm exp} \approx p_{\rm exp}^{\rm signal} + p_{\rm exp}^{\rm dark}
\ .
\end{equation}

The two contributions to the detected signals contribute differently
to the error rate.  The signal contributes an error with a probability
$p_e^{\rm signal}$,
per (arriving) signal,
due to
misalignment or polarization diffusion. The corresponding 
error probability 
per time slot is given by 
$e^{\rm signal} = p_{\rm exp}^{\rm signal} p_e^{\rm signal}$. 
On the other hand, a dark count contributes an error with probability 
$p_e^{\rm dark} = 1/2$, hence 
the corresponding contribution to the error rate 
per time slot is given by 
$ e^{\rm dark} = p_{\rm exp}^{\rm dark} (1/2)$.
The total error rate 
per time slot, $e$, is
$ e \approx p_{\rm exp}^{\rm signal} p_e^{\rm signal} +  
\frac{1}{2} p_{\rm exp}^{\rm dark}$, 
ignoring the coincidence term.
The contribution to the error rate per sifted
key bit is then given by  $ p_e = e/ p_{\rm exp}$. 
Let us consider the dark counts (in the absence of any signal)
to be a fixed parameter of an experimental setup.  
Thus, the dark count contribution to the error-rate becomes dominant when 
$p_{\rm exp}^{\rm signal}$ is
decreased, that is, when the distance is increased. 
In the following,
we use the bound 
$e >\approx \frac{1}{2} p_{\rm exp}^{\rm dark}$,
to obtain limits on the distance of secure \QKD.

If 
the error rate per sifted key bit $ p_e$ exceeds
$1/4$ and we (conservatively) 
assume that the eavesdropper has full control on the errors, there is no
way to create a secure key.
With such an allowed error-rate, 
a  simple intercept/resend attack causes Bob and
Eve to share (approximately) half of Alice's bits and to 
know nothing about the other half;
hence, Bob does not possess 
information which is unavailable to Eve, and no secret key can be distilled.
Using~$p_e = e/ p_{\rm exp}$ and $p_e < \frac{1}{4}$,
we obtain the necessary condition for secure \QKD\ 
\begin{equation}
\label{errorcriterion}
p_{\rm exp}> \frac{e}{1/4} \ ,
\end{equation}
and using Eq.~\ref{p-exp} and
$e \geq p_{\rm exp}^{\rm dark}/2$, we finally obtain
$p_{\rm exp}^{\rm signal} + p_{\rm exp}^{\rm dark} 
\geq 2 p_{\rm exp}^{\rm dark}$.   

For ideal single-photon states we therefore obtain (with $p_{\rm exp}^{\rm signal}= F \eta_{\rm B}$ and $p_{\rm exp}^{\rm dark} = d_{\rm B}$) 
the bound
$F \eta_{\rm B} + d_{\rm B} > 2 d_{\rm B} $ which finally gives
$F \eta_{\rm B} >  d_{\rm B} $.
We see 
that even for ideal single-photon sources (SP), the existence of a
dark count rate leads to a minimum transmission rate 
\begin{equation}
F > F_{\rm SP} = \frac{d_{\rm B}}{\eta_{\rm B}}
\end{equation}
below 
which \QKD\ cannot be securely implemented.
[Even for perfect detection 
efficiency ($\eta_{\rm B}=1$) we get a bound
$F > F_{\rm SP} = d_{\rm B}$.]
These bounds correspond, according to 
Eq.~\ref{distance}, to a maximal covered distance.

In a quantum optical implementation single-photon states would be
ideally suited for quantum key distribution.
However, such states have not yet been practically implemented for \QKD,
although proposals exist and experiments have been performed
to generate such states.
The signals produced in the experiments usually 
contain
zero, one, or more than one photon in the same polarization (with probabilities $p_0$ $p_1$,
and $p_{\rm multi}$ respectively.)  The multi-photon part of the signals
leads to a severe security hole, as has been anticipated earlier
\cite{BBBSS,HIGM,Yuen}.  Let us present the {\em photon number splitting }
(PNS) attack, which is a modification of an attack suggested
in~\cite{HIGM} (the attack of~\cite{HIGM} was disputed in~\cite{Yuen} 
so the modification
is necessary): Eve deterministically splits
one photon off each
multi-photon signal. To do so, she
projects the state onto subspaces characterised by $n$, 
which is the total photon number,
which can be measurement via a quantum nondemolition (QND) measurement.
The projection into these subspaces does not modify the polarization of the
photons.  Then she performs a polarization-preserving
splitting operation, for example, by an interaction described by a Jaynes-Cummings Hamiltonian
~\cite{moelmer} or an active arrangement of beamsplitters combined with further QND 
measurements.  She keeps one photon and sends the
other ($n-1$)~photons to Bob. We assume (conservatively)
that Bob's detector cannot resolve
the photon number of arriving signals.  When receiving the
data regarding the basis, Eve measures her photon and obtains full
information.
Each signal containing more than one photon in this way will yield
its complete information to an eavesdropper.

The situation becomes worse in
the presence of loss, in which case the eavesdropper can replace the
lossy channel by a perfect quantum channel and forward to Bob only
a chosen signal.
This suppression is controlled such that
Bob will find
precisely the number of signals as expected given the characterisation
of the lossy channel. If there is a strong contribution by multi-photon
signals, then
Eve can use only those signals and suppress the single-photon signals
completely, to obtain full information on the transmitted bits. 
Even  for  the case of perfect detectors in Bob's hands 
($\eta_{\rm B}=1$ and $d_{\rm B}=0$)
the above argument leads to a neccesary condition for security
\begin{equation} \label{multi-criterion}
 p_{\rm exp}^{\rm signal} > p_{\rm multi} .
\end{equation}
If this condition is violated, Eve gets full information without
inducing any errors. 
For given probabilities $p_1$ and $p_{\rm multi}$ (and given transmission rate $F$),
a bound on the distance is obtained, even for perfect detection.

We assume that Eve has control on $\eta_{\rm B}$ (which is 
a reasonable assumption, but somewhat conservative); also we use the standard
conservative 
assumption that all errors are controlled by Eve (including dark counts).
Without these assumptions, one gets much better security but one that is more
difficult to justify properly.

Whereas this work concentrates mainly on insecurity results, 
we also obtained here a result important for positive security
proofs:
For a general source (emitting into the four BB84 polarization modes) 
analyzing all possible
attacks in a large Hilbert space (the Fock space)
is a very difficult task.
However, if Alice can dephase the states to create a mixture of 
``number states'' (in the chosen BB84 polarization state)
the transmitted signals are replaced by mixed states. Then, these
states do not change at
all when Eve performs the QND part of the PNS attack!
[A QND measurement on the total photon
number].
Therefore Eve can start her attack by a PNS attack 
{\em without loss of generality}, and hence,
the PNS attack becomes the optimal attack.
Fortunately, in realistic scenarios the dephasing happens automatically
(or with little help of Alice)
Following this 
observation, a complete positive security proof against all individual particle
attacks has been subsequently given~\cite{nl99b}.
More sophisticated  
collective and joint attacks can also be restricted 
to the PNS attacks.

Let us return to the necessary condition for security. 
We can combine 
the idea of the two criteria above,
(\ref{errorcriterion}, \ref{multi-criterion}),
to a single, stronger one, given by
\begin{equation}
\label{combicriterion}
\frac{e}{1/4} < p_{\rm exp}-  p_{\rm multi} \; .
\end{equation}
This criterion 
stems from the scenario that Eve splits all
multi-photon signals while she  eavesdrops on some of the single-photon
signals. We can think of the key as of consisting of two parts: an
error free part steming from multi-photon signals, and a part with
errors coming from single-photon signals. The error rate in the second
key has therfore to obey the same inequality as used in criterion
(\ref{errorcriterion}).

We now 
explore the consequences of the necessary condition for
security for two practical
signal sources. These are the weak coherent pulses
and the signals generated by parametric downconversion.

In \QKD\ experiments the signal states are, typically, weak coherent
pulses containing, on average, much
less than one  photon. The information is contained in
polarization mode of the \WCP. 

Coherent states 
$|\alpha \rangle = e^{-|\alpha|^2 /2} \sum_n \alpha^n/{\sqrt n!}|n\rangle$ 
with amplitude $\alpha$ (chosen to be  
real) give a photon number 
distribution (per pulse~\cite{pulses})
$p_n  = e^{-\alpha^2} \left(\alpha^2
\right)^n/{n!}$.
Since we analyze PNS attack only, it doesn't matter if the realistic
``coherent state'' is a mixture of number states. 
Thus, 
$ p_{\rm exp}^{\rm signal} = \sum_{n=1}^\infty \exp(-F \eta_{\rm B} \alpha ^2) \left(F
\eta_{\rm B} \alpha^2 \right)^n/{n!}$ and
$p_{\rm multi}  =  \sum_{n=2}^\infty \exp(-\alpha^2) \left(\alpha^2
\right)^n/{n!} $.  
With $p_{\rm exp} \approx p_{\rm exp}^{\rm signal}+ d_{\rm B}$ and the error rate 
$e=\frac{1}{2} d_{\rm B}$ we find for $\alpha^2 \ll 1$
(by expanding to 4th oder in $\alpha$ and neglecting the term
proportional to $F^2 \eta_{\rm B}^2 \alpha^4$) the
result
\begin{equation}
F >  \frac{d_{\rm B}}{\eta_{\rm B} \;\alpha^2}+\frac{\alpha^2}{\eta_{\rm B} \;2}  \; .
\end{equation}
Obviously, there is an optimal choice for $\alpha^2$
which leads to the bound
\begin{equation}
F>F_{\rm WCP} = \sqrt{2 d_{\rm B}} / \eta_{\rm B} .
\end{equation}
To illustrate this example we insert numbers taken from the experiment
performed at $1.3 \mu m$ by Marand and Townsend \cite{marand95a}. 
We 
use $\eta_{\rm B} = 0.11$, $d_{\rm B} = 10^{-5}$. Then the criterion gives $F >
0.041$. With a constant loss of $5$ dB and a fiber loss at $0.38$
dB/km, this is equivalent, according to Eq.~(\ref{distance}), 
to a maximum distance of $24$ km at an
average (much lower than standard) 
photon number of $4.5 \, \times\, 10^{-3}$. As we used
approximations in the evaluation of Eq.~(\ref{combicriterion}),
the achievable distance could differ slightly from this values either way.

With $\alpha^2 = 0.1$, as in the literature,
secure transmission to any distance is impossible, according to our conditions.
In that case, even if we assume $\eta_{\rm B}$ to be out of control of the
eavesdropper, we find that secure transmission to distance of 
$21$ km is impossible.
Frequently we find even higher average photon numbers in
the literature.

The \WCP\ scheme seems to be prone to difficulties due to the high
probability of signals carrying no photons (the vacuum contribution).
This can be overcome in
part by the use of parametric downconversion (\PDC) scheme, or other single photon
sources. \PDC\ has
been used before for \QKD\ \cite{PDCref}. 
We use a
different formulation which enable us to
analyse the advantages and limits of the \PDC\ method relative to the \WCP\
one.

To approximate a single-photon state we use a
\PDC{} process where we
create the state in an output  mode described by photon creation
operator $a^\dagger$ conditioned on the detection of a photon in
another mode described by $b^\dagger$.  If we neglect dispersion, then
the output of the \PDC{} process is described \cite{Walls94} on the two
modes with creation operators $a^\dagger$ and $b^\dagger$ using the
operator $T_{a\,b}(\chi) = \exp\left\{{\rm i} \chi(a^\dagger b^\dagger - a b)
\right\}$, 
with $\chi \ll 1$, as
$
|\Psi_{a\,b}\rangle  =  T_{a\,b}(\chi) |0,0\rangle  \approx 
 (1- \chi^2/2 + \frac{5}{24} \chi^4)|0,0\rangle +
(\chi - \frac{5}{6} \chi^3) |1,1\rangle 
 + (\chi^2 - \frac{7}{6} \chi^4)|2,2\rangle + \chi^3 |3,3\rangle +
\chi^4 |4,4\rangle
$. 
The states in this description are states of photon flux,
and we assume the addition
of choppers in order to deal with photon numbers per pulse as in 
the \WCP{} case.

If we had an ideal detector resolving photon numbers 
(that is, a perfect counter)
then we could
create a perfect single-photon state by using the state in mode $a$
conditioned on the detection of precisely one photon in the pulse in
mode $b$. However, realistic detectors useful for this task have a
single-photon detection efficiency far from unity and can resolve the
photon number only at high cost, if at all. Therefore, we assume a
detection model which is described by a finite detection efficiency
$\eta_{\rm A}$ and gives only two possible outcomes: either it is not
triggered or it is triggered, thereby showing that at least one photon
was present. The detector may experiences a dark count rate at $d_{\rm A}$
per time slot.  The two POVM elements describing this kind of detector
can be approximated for our purpose  by
$E_0 = (1-d_{\rm A})|0\ra\la 0| + 
\sum_{n=1}^{\infty} (1-\eta_{\rm A})^n |n\ra\la n| $
and
$E_{\rm click}= d_{\rm A} |0\ra\la 0| + \sum_{n=1}^{\infty}
(1-(1-\eta_{\rm A})^n) |n\ra\la n| $.
The reduced density matrix for the output signal in mode $b$ conditioned on a
click of the detector monitoring mode $a$ is then given by
\begin{eqnarray}
\label{PDCsignal}
\rho &=& \frac{1}{N} {\rm Tr}_b\left[|\Psi_{a \,b}\ra\la \Psi_{a \,b}|
 E_{\rm click}\right]  \approx  \frac{1}{N}
\Big[ d_{\rm A} (1-\chi^2+\frac{2}{3} \chi^4)
 |0\ra\la 0| \nonumber \\
 &+& \eta_{\rm A} \chi^2 (1-\frac{5}{3} \chi^2) |1\ra\la 1| 
  +
 \eta_{\rm A} (2- \eta_{\rm A}) \chi^4 |2\ra\la 2| \Big] 
\end{eqnarray}
with the normalization constant $N$.  
To create the four signal states we rotate the polarization
of the signal, for example using a beam-splitter and a phase shifter.
Note that a mixture of Fock states is created by the detection process, 
so that the PNS attack is optimal for Eve.

After some calculation following the corresponding calculation in the 
\WCP\ case, the necessary condition for security
(\ref{combicriterion}) takes for the signal state (\ref{PDCsignal}) the
form
\begin{equation}
F > \frac{d_{\rm A}\; d_{\rm B}}{\eta_{\rm A} \;\eta_{\rm B}\; \chi^2} + \frac{d_{\rm B}}{\eta_{\rm B}}
+ \frac{2-\eta_{\rm A}}{\eta_{\rm B}} \chi^2
\end{equation}
since we assume $d_{\rm B} \ll 1$ and $\chi^2 \ll 1$ and neglect terms going
as $\chi^4$, $d_{\rm B} d_{\rm A}$, and $\chi^2 d_{\rm B}$.
The first error term is due to coincidence of dark counts, the second error
term is due to
coincidence of a photon loss and a dark count at Bob's site; the third
term is the effect of multi photon signal (signals that leak full
information to the eavesdropper).
As we did in the \WCP\ case, we optimize this expression to find the
criterion
\begin{equation}
F>F_{\rm PDC} = 2 \sqrt{\frac{d_{\rm A}\; d_{\rm B}\; (2-\eta_{\rm A})}{\eta_{\rm A}\; \eta_{\rm B}^2}}
+ \frac{d_{\rm B}}{\eta_{\rm B}} \; .
\end{equation}
If we now assume that Alice and Bob use the same detectors as in the
\WCP\ case with the numbers provided by \cite{marand95a} we obtain
$F_{\rm PDC} > 8.4 \times 10^{-4}$ corresponding via Eq.~(\ref{distance})
to a length of $68$ km.

Since we can use down-conversion set-ups which give photon pairs with
different wavelength, we can use sources so that one photon has the
right wavelength for transmission over long distances, e.g.~1.3
$\mu$m, while the other photon has a frequency which makes it easier
to use efficient detectors \cite{PDCref}.
In the limit of Alice using perfect detectors (but not perfect counters)
($\eta_{\rm A}=1$ and $d_{\rm A}=0$), we obtain
\begin{equation} F_{{\rm PDC}} = d_{\rm B} /\eta_{\rm B} 
\ ,
\end{equation}
as for the single-photon source.
leading to a maximal distance of $93$ km.
This optimal distance
might also be achievable using new single-photon sources
of the type suggested in~\cite{kim99a}.

We have shown a necessary condition for secure \QKD\ which uses
currently experimental implementations. We find that secure \QKD\
might
be achieved with the presented experiments using \WCP\ if one would use
appropriate parameters for the expected photon number, which are
considerably lower than those used today. 
With current parameters, it seems that all current
experiments cannot be proven secure.
The distance which can
be covered by \QKD\ is mainly limited by the fiber loss, 
but \WCP\ schemes might be
totally insecure even to zero distance (in several of the existing experiments),
due to imperfect detection.
The distance can be increased by the use of parametric downconversion
as a signal source, but even in this case the fundamental
limitation of the range 
persists, and a radical reduction of $\beta$ or of the dark
counts is  required in order to increase the distance 
to thousands of kilometers.

The use of quantum repeaters
(based on quantum 
error-correction or entanglement purification) in the far future, can 
yield secure transmission to any distance, and the security is not
altered even if the repeaters are controlled by Eve~\cite{mor-thesis}.

Parts of this work were presented independently by G.B., T.M., and
B.C.S., and by N.L. during the QCM'98 conference.
We would like to thank C.~H.~Bennett, H.~Yuen, E.~Polzik and J.~H.~Kimble
for their helpful comments.
N.L. thanks the Academy of Finland for financial support. 
The work of T.M. was supported in part by grant \#961360 from the Jet
Propulsion Lab,
and grant \#530-1415-01 from the DARPA Ultra program.

\end{multicols}


\begin{thebibliography}{99}
\bibitem {BB84} C.\,H.~Bennett and G.~Brassard, in {\em Proc.\ of IEEE
    Inter.\ Conf.\ on Computers, Systems and Signal Processing,
    Bangalore, India} (IEEE, New York, 1984) p.~175.
\bibitem{BBE} C.\,H.~Bennett, G.~Brassard, and A.~Ekert,
 Scientific American, {\bf 257}, 50 (1992).
\bibitem{Shor97} P.~W.~Shor, 
 Siam J. of Comp., {\bf 26}, 1484 (1997).
\bibitem {BM97b} E.~Biham and T.~Mor,
 Phys.\ Rev.\ Lett.\ {\bf 79}, 4034 (1997);
\bibitem{mayers}
D. Mayers, {\it Proc. of Crypto'96, LNCS} {\bf 1109}, 343
(1996); D. Mayers, quant-ph/9802025.
\bibitem{nl99b} N. L\"utkenhaus, Acta Phys. Slovaca {\bf 49}, 549
(1999); quant-ph/9910093.
\bibitem{BBBSS} C.\,H.~Bennett, F.~Bessette, G.~Brassard, L.~Salvail
and J.~Smolin,
 J. Crypto.\ {\bf 5}, 1 (1992).

\bibitem{bennett95a} 
C.~H. Bennett, G. Brassard, C. Cr\'epeau, and U.~M. Maurer,
 IEEE Trans. Inf. Theo. {\bf 41,} 1915 (1995).

\bibitem{WCPexperiments} J.~D.~Franson and H.~Ilves, J. of Modern Optics
{\bf 41,} 2391 (1994);
P.~D. Townsend,   IEEE
  Photonics Technology Letters {\bf 10,} 1048 (1998);
W.~T. Buttler, R.~J. Hughes, P.~G. Kwiat, G.~G. Luther, G.~L. Morgan, J.~E.
  Nordholt, C.~G. Peterson, and C.~M. Simmons,   Phys. Rev. A {\bf 57,} 2379 (1998).

\bibitem{HIGM} B.~Huttner, N.~Imoto, N.~Gisin, and T.~Mor,
 Phys. Rev. A. {\bf 51}, 1863 (1995).

\bibitem {Yuen} H.~P.~Yuen,  Quant.\ Semiclass.\ Opt.\ {\bf 8},
939 (1996).
\bibitem{moelmer} Klaus M\o lmer, private communication.
\bibitem{marand95a}
C. Marand and P.~T. Townsend,  Opt. Lett. {\bf 20,} 1695 (1995).
\bibitem {PDCref} A.\,K.~Ekert, J.\,G.~Rarity, P.\,R.~Tapster and G.\,M.~Palma,    Phys.\ Rev.\ Lett.\ {\bf 69 }, 1293 (1992);
\bibitem {Walls94} D.\ F.\ Walls and G.\ J.\ Milburn,
        {\em Quantum Optics} (Springer-Verlag, Heidelberg, 1994).
\bibitem{pulses} K.\,J.~Blow, R.~Loudon, S.\,J.\,D.~Phoenix and T.\,J.~Sheperd,
     Phys.\ Rev.\ A {\bf 42}, 4102 (1990).

\bibitem{kim99a}
J.~Kim, O.~Benson, H.~Kan, and Y.~Yamamoto,  Nature {\bf 397},
500 (1999).

\bibitem{mor-thesis} T.~Mor, Ph.D.~Thesis, Technion, Haifa (1997);
quant-ph/9906073. 

\end{thebibliography}
\end{document}